\documentclass[a4paper,oneside,12pt]{article}
\usepackage[english]{babel}
\usepackage{amsmath}
\usepackage{amsfonts}
\usepackage{amsthm}

\theoremstyle{plain} \newtheorem{theorem}{Theorem}
\theoremstyle{remark} \newtheorem*{remark}{Remark}
\theoremstyle{plain} \newtheorem{lemm}{Lemm}
\everymath{\displaystyle}
\numberwithin{equation}{section}

\begin{document}

\LARGE \noindent\textbf{Inverse problems in the multidimensional \\ hyperbolic equation with rapidly oscillating \\ absolute term}

\normalsize Babich P.V., Levenshtam V.B.

\footnotesize \textbf{Abstract.} The paper is devoted to the development of the theory of inverse problems for evolution equations with terms rapidly oscillating in time. A new approach to setting such problems is developed for the case in which additional constraints are imposed only on several first terms of the asymptotics of the solution rather that on the whole solution. This approach is realized in the case of a multidimensional hyperbolic equation with unknown absolute term.



\section*{Introduction}

We consider some problems of recovering rapidly oscillating in time absolute term from certain data on a partial asymptotics of the solution. Hence we study some of the coefficient inverse problems. The theory of inverse problems was the subject of many monographs (see, e.g. \cite{bib17}--\cite{bib18}) and papers (see, e.g. \cite{bib17}--\cite{bib18}). But there are almost no problems with rapidly oscillating data in the classical theory of inverse problems.

This paper as paper \cite{bib211} was motivated by the paper \cite{bib18},  in which inverse problems for the one-dimensional wave equation with unknown absolute term
was posed and solved. In \cite{bib18} right-hand side represented in the form $f(x)r(t)$, where $r$ is unknown. An additional condition in \cite{bib18} was the value of $q(t)$ of the solution at a fixed point $x = x_0$. In \cite{bib211} we have the same form of right-hand side of multidimensional hyperbolic equation, but the unknown term rapidly oscillate: $r = r(t,\omega t), \omega \gg 1$. This brings up the question, should we impose an additional condition on the whole solution, as in \cite{bib18}. In paper \cite{bib211} it was established that the additional condition may be imposed only on several first coefficients of the asymptotics of the solution rather than on the whole solution. In the present paper following inverse problem are solved:
1) $f$ is unknown;
2) $f$ and fast component of $r$ are unknown.

In conclusion, we mention that, problems with data rapidly oscillating in time model many physical (and other) processes (in particular, related to high-frequency mechanical, electromagnetic, and other actions on a medium) see, for example, \cite{bib19}--\cite{bib22}. The inverse problems with such specificity have been studied in \cite{bib211}, \cite{bib1}, \cite{bib2} by us.

\section{Principal symbols}
\label{defenit}

Let $\Omega$ denote a bounded domain in $\mathbb{R}^n, n \in \mathbb{N}. S$ its boundary. We denote the open cylinder $\Omega \times (0,T) \subset \mathbb{R}^{n+1}$ by $Q_T$, its closure  $\overline{Q}_T$.
Consider the following hyperbolic initial boundary-value problem with a large parameter $\omega$:
\begin{equation}
\frac{\partial^2 u}{\partial t^2} =  L u + f(x,t) r(t, \omega t), (x,t) \in Q_T,
\label{s1}
\end{equation}
\begin{equation}
\left. u \right|_{t = 0} = 0, \; \left. \frac{\partial u}{\partial t} \right|_{t = 0} = 0,
\label{s2}
\end{equation}
\begin{equation}
\left. u \right|_{x \in S} = 0,
\label{s3}
\end{equation}
All functions are real. We consider that the symmetric differential expression
\begin{equation}
Lu = \sum\limits_{i,j = 1}^n \frac{\partial}{\partial x_i} \left[ a_{ij} (x) \frac{\partial u}{\partial x_j} \right] - c(x) u \, -
\label{eqlu}
\end{equation}
is defined in $\Omega$ and satisfies the ellipticity condition, so that
\begin{equation}
a_{ij} (x) = a_{ji} (x), \sum\limits_{i,j = 1}^n a_{ij} (x) \xi_i \xi_j \geq \gamma \sum\limits_{i = 1}^n \xi_i^2, \text{where } \gamma = const > 0,
\label{aij}
\end{equation}
for all $x \in \Omega$ and any real vector $\xi = (\xi_1, \xi_2, ... , \xi_n)$.

We shall assume that the function $r(t,\tau)$ is defined and is continuous on the set $D = \{ (t,\tau): (t,\tau) \in [0,T]\times [0,\infty) \}$ and $2\pi$-periodic in $\tau$. Let us represent it as the sum:
\begin{equation*}
r (t,\tau) = r_0 (t) + r_1 (t,\tau),
\end{equation*}
where $r_0 (t)$ -- is the mean value of $r(t,\tau)$ over $\tau$:
\begin{equation*}
r_0 (t) = \left\langle r (t,\cdot) \right\rangle = \left\langle r (t,\tau) \right\rangle_{\tau} \equiv \frac{1}{2\pi} \int\limits_0^{2\pi} r (t,\tau) d\tau.
\label{r_1}
\end{equation*}

\section{The auxilary results}
\subsection{The results of V.A. Il'in \cite{bib23}}

Lets consider the problem
\begin{equation}
\frac{\partial^2 u}{\partial t^2} =  L u + F(x,t), (x,t) \in Q_T,
\label{ss1}
\end{equation}
\begin{equation}
\left. u \right|_{t = 0} = \varphi (x), \; \left. \frac{\partial u}{\partial t} \right|_{t = 0} = \psi (x),
\label{ss2}
\end{equation}
\begin{equation}
\left. u \right|_{x \in S} = 0,
\label{ss3}
\end{equation}
Let domain $\Omega$, the coefficients of the expression $L$ \eqref{eqlu}, right-hand side $F$ and initial conditions $\varphi$ è $\psi$ satisfy the following conditions.

I. $\Omega$ is bounded connected domain in $\mathbb{R}^n, n \in \mathbb{N}$, contained, together with its boundary $S$, in an open domain $C \in \mathbb{R}^n$. \footnote{Recall that a domain is said to be normal if the Dirichlet problem for the Laplace equation in this domain is solvable for continuous boundary function.}

II. Coefficients $a_{ij} (x)$ and $c(x)$ ensure existence of full orthonormal in $L_2 (\Omega)$ system classic eigenfunctions of problem
\begin{equation*}
\left\{
\begin{array}{l}
Lu = \lambda u, \\
u|_{S} = 0.
\end{array}
\right.
\end{equation*}
To do this, since \cite{bib23} it suffices to provide further conditions. Functions $a_{ij} (x), c (x)$ can be continued to domain $C$ so that $a_{ij} \in C^{1+\mu} (C), c \in C^{\mu} (C), \mu \geq 0. $
Moreover, $a_{ij} \in C^{\left[ \frac{n}{2} \right] + 2} (\overline{\Omega}), c \in C^{\left[ \frac{n}{2} \right] + 1} (\overline{\Omega})$. Let $y_m, \lambda_m, m = 1,2,...,$ denote eigenfunctions and eigenvalues noted above. We shall assume that $\{\lambda_m\}$ is nondecreasing sequence: $0 < \lambda_1 \leq \lambda_2 \leq ...$

III. Initial functions $\varphi \in C^{\left[ \frac{n}{2} \right] + 3} (\overline{\Omega}), \psi \in C^{\left[ \frac{n}{2} \right] + 2} (\overline{\Omega})$ and $\left. \varphi \right|_{x \in S} = \left. L \varphi \right|_{x \in S} = ... = \left. L^{ \left[ \frac{n+4}{4} \right]} \varphi \right|_{x \in S} = 0, \left. \psi \right|_{x \in S} = \left. L \psi \right|_{x \in S} = ... = \left. L^{ \left[ \frac{n+2}{4} \right]} \psi \right|_{s \in S} = 0$. Let $\varphi_m, \psi_m$ denote the coefficients of the Fouries expansion of functions $\varphi (x), \psi(x)$ in the basis of $y_m$.

IV. The right-hand side $F \in C([0,T], C^{\left[ \frac{n}{2} \right] + 2} (\overline{\Omega})), \left. F \right|_{x \in S} = \left. L f \right|_{x \in S} = ... = \left. L^{\left[\frac{n+2}{4}\right]} f \right|_{x \in S} = 0.$ Let $F_m (t)$ denote the coefficients of the Fouries expansion of function $F (x,t)$ in the basis of $y_m$

\begin{theorem}
(V.A. Il'in) If conditions I--IV hold, the series
\begin{equation}
u (x,t) = \sum\limits_{m=1}^{\infty} y_m (x) \left[ \varphi_m \cos{\sqrt{\lambda_m} t} + \frac{\psi_m}{\sqrt{\lambda_m}} \sin{\sqrt{\lambda_m}t} \right] + \sum\limits_{m=1}^{\infty} y_m (x) \frac{1}{\sqrt{\lambda_m}} \int\limits_0^t F_m (\tau) \sin{\sqrt{\lambda_m}(t-\tau)} d\tau
\label{supp1}
\end{equation}
and the series $u_t, u_{tt}$ obtained by single and double differentiation of \eqref{supp1} with respect to $t$ are converge uniformly in $\overline{Q_T}$. The series $u_{x_i}, u_{tx_i}, u_{x_i x_j}$ obtained by single and double differentiation of \eqref{supp1} with respect to any two variables are converge uniformly in any domain that is  strictly contained in $Q_T$. At the same time, \eqref{supp1} is classic solution of \eqref{ss1}-\eqref{ss3}.
\label{thilin}
\end{theorem}

This result can be found in \cite[òåîðåìû 6, 8]{bib23}\footnote{Here and in what follows, we use results of \cite{bib23} in "classical" terms (see \cite[Remark 3, p. 114 of the Russian original]{bib23}). In \cite{bib23}, such classical versions are not stated explicitly, but when referring to results of \cite{bib23}, we always mean their classical versions.}

\begin{lemm}
If conditions I, II, III hold, bilinear series for eigenfunctions
$\sum\limits_{m = 1}^{\infty} \frac{y_m^2 (x)}{\lambda_m^{\left[\frac{n}{2}\right] + 1}}$
is converges uniformly in
$\overline{\Omega}$,
bilinear series
$\sum\limits_{m = 1}^{\infty} \frac{\left| \frac{\partial v_m (x)}{\partial x_i} \right|^2}{\lambda_m^{\left[\frac{n}{2}\right]+2}}$
and
$\sum\limits_{m = 1}^{\infty} \frac{\left| \frac{\partial^2 y_m (x)}{\partial x_i \partial x_j} \right|^2}{\lambda_m^{\left[\frac{n}{2}\right]+3}}$
are converge uniformly in any domain that is  strictly contained in $\Omega' \subset \Omega$.
\label{ilinlemm0}
\end{lemm}

\begin{lemm}
Let coefficients $a_{ij} (x)$ continuous there together with their derivatives up to order $k$, and $c(x)$ continuous there together with its derivatives up to order $k-1$. We shall assume that function $\Phi (x), x \in \overline{\Omega}$ satisfies following conditions:

1) $\Phi \in C^{k+1} (\overline{\Omega})$,

2) $ \left. \Phi \right|_{x \in S} = \left. L \Phi \right|_{x \in S} = ... = \left. L^{\left[\frac{k}{2}\right]} \Phi \right|_{x \in S} = 0.$

Then for $\Phi$ inequality of Bessel type holds true:
\begin{equation*}
\sum\limits_{m = 1}^{\infty} \Phi_m^2 \lambda_m^{k+1} \leq
\left\{
\begin{array}{l}
\int\limits_{\Omega} \left[ \sum_{i,j=1}^n a_{ij} \frac{\partial}{\partial x_i} (L^{\frac{k}{2}} \Phi) \frac{\partial}{\partial x_j} (L^{\frac{k}{2}} \Phi) + c (L^{\frac{k}{2}} \Phi)^2 \right] dx, k -- \text{÷åòíîå}, \\
\int\limits_{\Omega} \left[ L^{\frac{k+1}{2}} \Phi \right]^2 dx, k - \text{íå÷åòíîå}.
\end{array}
\right.
\end{equation*}
\label{ilinlemm}
\end{lemm}

\subsection{The problem 1}
\label{dirprob1}
\textbf{The direct problem 1. The three-term asymptotics}

Consider problem \eqref{s1}-\eqref{s3}, where domain $\Omega$, elliptic differential expression $L$ are the same as in Theorem \ref{thilin}.

Concerning the function $f(x,t)$ defined at $(x,t) \in \overline{Q}_T$, we assume that there exist continuous functions $f, Lf, f_t, f_{tt}, f_{ttt}$ and $Lf_t$, such that all of them belong to the space of functions $C_{t,x}^{0, \left[\frac{n}{2}\right]+2} (\overline{Q}_T)$ and, moreover,
$$\left.f\right|_{x \in S} = \left. L f\right|_{x \in S} = ... = \left. L^{\left[ \frac{n+6}{4} \right]} f\right|_{x \in S} = 0.$$
For brevity, we refer to functions r with these properties as functions of class $\mathbf{F_1}$.

We shall assume that the function $r(t,\tau)$ is defined and is continuous on the set $D = \{ (t,\tau): (t,\tau) \in [0,T]\times [0,\infty) \}$ and $2\pi$-periodic in $\tau$. As in Sec.\ref{defenit} let us represent $r$ as the sum of slow and oscillating components:
\begin{equation*}
r (t,\tau) = r_0 (t) + r_1 (t,\tau);
\end{equation*}
we shall assume that $r_0 \in C ([0,T])$, and the functions $r_1, r_{1t}, r_{1tt}$, and $r_{1ttt}$ belong to the class $ C (D).$ We denote function $r$ with such properties as function of class $\mathbf{R_1}$.

In the present paper, by a solution of problem \eqref{s1}-\eqref{s3} we mean its classical solution, i.e., a  function $u \in C(\overline{Q}_T)$, which has continuous derivatives $u_t \in C(\overline{Q}_T), u_{tt}$, and $u_{x_i x_j} \in C (Q_T), i,j = \overline{1,n},$ and satisfies relations \eqref{s1}--\eqref{s3}. Under our assumptions, the solution
of problem \eqref{s1}--\eqref{s3}, exists and is unique according to the Theorem 1.

Below we define functions and constants needed in what follows:
\begin{multline}
\rho_0 (t,\tau) = \int\limits_0^{\tau} \left( \int\limits_0^{p} r_1 (t,s) ds - \left\langle \int\limits_0^{\tau} r_1 (t,s) ds \right\rangle_\tau \right) dp - \\
     \left\langle \int\limits_0^{\tau} \left( \int\limits_0^{p} r_1 (t,s) ds - \left\langle \int\limits_0^{\tau} r_1 (t,s) ds \right\rangle_\tau \right) dp \right\rangle_{\tau}
\label{rho}
\end{multline}
\begin{equation*}
\rho_1 (t,\tau) = \left\langle \int\limits_0^{\tau} \rho_0 (t,s) ds \right\rangle_{\tau} - \int\limits_0^{\tau} \rho_0 (t,s) ds.
\end{equation*}
\begin{equation}
b_{1,m} = -  \rho_{0\tau} (0,0) f_m (0),
\label{b1}
\end{equation}
\begin{equation}
d_m = - \rho_0 (0,0) f_m (0),
\label{dm}
\end{equation}
\begin{equation}
b_{2,m} = - (2 \rho_1 (0,0) + \rho_0 (0,0)) f'_m (0) - ( 2 \rho_{1t} (0,0) + \rho_{0t} (0,0)) f_m (0),
\label{b2}
\end{equation}
where the $f_m (t)$ are the coefficients of the Fouries expansion of $f(x,t)$ in the basis of $y_m$.

Let us represent the solution of problem (\ref{s1})-(\ref{s3}) in the form:
\begin{equation}
u_{\omega} (x,t) = U_{\omega} (x,t) + W_{\omega} (x,t), \omega \gg 1,
\label{uw}
\end{equation}
\begin{equation}
U_{\omega} (x,t) = u_0 (x,t) + \omega^{-1} u_1 (x,t) + \omega^{-2} \bigl[ u_2 (x,t) + v_2 (x,t,\omega t)\bigr], \omega \gg 1,
\label{uww}
\end{equation}
\begin{equation}
u_0 (x,t) =  \sum_{m = 1}^{\infty} \frac{y_m (x)}{\sqrt{\lambda_m}} \int\limits_0^t f_m (s) r_0 (s) \sin{\sqrt{\lambda_m} (t-s)} ds,
\label{u_0}
\end{equation}
\begin{equation}
u_1 (x,t) = \sum_{m = 1}^{\infty} \frac{b_{1,m}}{\sqrt{\lambda_m}} y_m (x) \sin{\sqrt{\lambda_m} t},
\label{u_1}
\end{equation}
\begin{equation}
v_2 (x,t,\tau) = f(x,t) \rho_0 (t,\tau),
\label{v_2}
\end{equation}
\begin{equation}
u_2 (x,t) = \sum_{m = 1}^{\infty} y_m (x) \left( d_m \cos{\sqrt{\lambda_m} t} + \frac{b_{2,m}}{\sqrt{\lambda_m}} \sin{\sqrt{\lambda_m} t} \right).
\label{u_2}
\end{equation}
Note that, in view of the Theorem \ref{thilin}, the series \eqref{u_0}--\eqref{u_2} converge uniformly and absolutely.

\begin{theorem}
The solution $u_{\omega}(x,t)$ of problem (\ref{s1})-(\ref{s3}) can be expressed in the form (\ref{uw})-(\ref{u_2}), where
\begin{equation}
\bigl\| W_{\omega} (x,t) \bigr\|_{C(\overline{Q}_T)} = o(\omega^{-2}), \omega \to \infty. \label{th1}
\end{equation} \label{theor1}
\end{theorem}

\textbf{The inverse problem 1}
\label{sect3}

Suppose that the function $f(x, t)$ in the initial boundary-value problem \eqref{s1}-\eqref{s3} is the function of class $\mathbf{F_1}$ and the function $r \in \mathbf{R_1}$ is unknown. Choose a point $x^0 \in \Omega$ at which $f(x^0,t) \neq 0, t \in [0,T]$, and functions $\varphi_0 (t)$ and $\chi(t,\tau)$ satisfying the conditions:
\begin{eqnarray*}
\varphi_0 \in C^1 ([0,T]), \; \varphi_0 (0) = 0, \; \varphi'_0 (0) = 0;\label{eq19'} \\
\chi \in C^{3,2} (D),
\end{eqnarray*}
where the function $\chi(t,\tau)$ is $2\pi$-periodic in $\tau$ and has zero mean $\bigl( \left\langle \chi (t,\cdot) \right\rangle = 0 \bigr)$. Consider the functions $\varphi_1(t)$ and $\varphi_2(t)$ defined by
\begin{equation}
\varphi_1 (t) = \sum_{m = 1}^{\infty} \frac{b_{1,m}}{\sqrt{\lambda_m}} y_m (x^0) \sin{\sqrt{\lambda_m} t},
\label{phi1}
\end{equation}
\begin{equation}
\varphi_2 (t) = \sum_{m = 1}^{\infty} y_m (x^0) \left( d_m \cos{\sqrt{\lambda_m} t} + \frac{b_{2,m}}{\sqrt{\lambda_m}} \sin{\sqrt{\lambda_m} t} \right),
\label{phi2}
\end{equation}
where the $b_{1,m}, b_{2,m}$ and $d_{m}$ are the same as in \eqref{b1}-\eqref{dm}, but $\rho_0 (t,\tau)$ is now defined by
\begin{multline*}
\rho_0 (t,\tau) = \frac{1}{f(x^0,t)} \Biggl( \int\limits_0^{\tau} \left( \int\limits_0^{p} \chi_{s s} (t,s) ds - \left\langle \int\limits_0^{\tau} \chi_{s s} (t,s) ds \right\rangle_\tau \right) dp - \\
     \left\langle \int\limits_0^{\tau} \left( \int\limits_0^{p} \chi_{s s} (t,s) ds - \left\langle \int\limits_0^{\tau} \chi_{s s} (t,s) ds \right\rangle_\tau \right) dp \right\rangle_{\tau} \Biggr).
\end{multline*}

The inverse problem 1 is to find a function $r \in \mathbf{R_1}$ for which the solution $u_{\omega} (x,t)$ of problem \eqref{s1}-\eqref{s3} satisfies the condition
\begin{equation}
\left\| u_{\omega} (x^0,t) - \left[ \varphi_0 (t) + \frac{1}{\omega}\varphi_1 (t) + \frac{1}{\omega^2} \bigl( \varphi_2 (t) + \chi (t,\omega t) \bigr) \right] \right\|_{C([0,T])} = o (\omega^{-2}), \; \omega \to \infty.
\label{babich-inv}
\end{equation}

\begin{theorem}
For any pair of functions $\chi, \varphi_0$ and point $x^0$ satisfying the conditions specified above inverse problem 1 is uniquely solvable.
\label{babich-t4}
\end{theorem}
\begin{remark}
Finding the function $r_0$ reduces to solving a Volterra equation of the second kind
\begin{equation}
f(x^0,t)r_0(t) + \int_0^t K(t,s) r_0(s) \, ds = \varphi_0'' (t),
\label{volt}
\end{equation}
\begin{equation*}
K(t,s) = - \sum\limits_{m = 1}^{\infty} \sqrt{\lambda_m} f_m (s) \sin{\left(\sqrt{\lambda_m}(t-s)\right)} y_m (x^{0}).
\end{equation*}
Function $r_1$ calculated by
\begin{equation}
r_1 (t,\tau) = \frac{1}{f(x^0,t)} \frac{\partial^2}{\partial \tau^2} \chi (t,\tau),
\label{r1}
\end{equation}
\end{remark}
\begin{remark}
The Theorem \ref{theor1} and \ref{babich-t4} can be found together with their proof in paper \cite{bib211}.
\end{remark}

\subsection{The lemm Krasnosel'skii et al. \cite[Sec. 22.1]{bib24}}

Suppose that $\Omega$ is bounded connected domain in $\mathbb{R}^n$ and $S$ its boundary. We denote $k_0 > \frac{n}{4}$ is natural value such that $S \in C^{2k_0}$ and functions $b_{ij}, d \in C^{2k_0 - 2} (\overline{\Omega})$. Moreover, boundary smoothness meant in the same manner as in \cite[Theorem 15.2]{bib29}. In space $L_2 (\Omega)$ consider elliptic differential operator
\begin{equation}
L_0 u = \sum\limits_{i,j = 1}^n b_{ij} (x) \frac{\partial^2 u}{\partial x_i \partial x_j} - d(x) u, \; u \in D(L_0) \equiv \dot{W}_2^2 (\Omega),
\end{equation}
where $\dot{W}_2^2 (\Omega)$ is closure in $W_2^2 (\Omega)$ of set of smooth finite in $\Omega$ function. We shall assume that coefficient $d(x)$ is so large that $L_0$ is invertible operator. Results of \cite{bib29} imply the estimate
\begin{equation}
\left\| L_0^{k_0} u \right\|_{L_2} \geq c \left\| u \right\|_{W_2^{2 k_0}} , \; u \in D(L_0^{k_0}), c -- \text{positive value}.
\end{equation}
We assume that the domain $\Omega$ satisfies Sobolev's imbedding Theorem:
\begin{equation}
\|u\|_{C^l (\overline{\Omega})} \leq c \| u \|_{W_2^s (\Omega)}, \; u \in W_2^s (\Omega),
\end{equation}
where $s-l > \frac{n}{2}, c $ is positive value. Classic condition for this Theorem is that $\Omega$ is star domain.

The above leads to the following result:
\begin{lemm}
For any integer $|r| \in [0, 2k_0 - \frac{n}{2}]$ operator $D^r L_0^{-k_0}$ continuously acts from $L_2 (\Omega)$ to $C^{2k_0-r-\frac{n}{2}} (\overline{\Omega})$, where $D^r u = \frac{\partial^r u}{\partial x_1^{r_1} ... x_n^{r_n}}, r = (r_1, ..., r_n)$ is multi-index with length $|r| = r_1 +...+r_n.$
\label{th5}
\end{lemm}
Lemm \ref{th5} can be found in \cite[ï.22.2]{bib24} without specialization of some requirements to coefficients and boundary.

\section{The main results}

\subsection{The problem 2}
\label{dirproblem}

\textbf{The direct problem 2. The main term of asymptotics}

Let as in Sec.\ref{dirprob1} $\Omega$ and operator $L$ satisfies Theorem \ref{thilin} conditions.

Let us consider the problem \eqref{s1}-\eqref{s3}. From this point onward function $f(x,t)$ is invariant with $t$ h.e. $f(x,t) = f(x), x \in \Omega$. We also assume that $f \in C^{\left[ \frac{n}{2} \right] + 2} (\overline{\Omega})$,
\begin{equation}
\left. f \right|_{x \in S} = \left. L f \right|_{x \in S} = \left. L^2 f \right|_{x \in S} = ... = \left. L^{\left[ \frac{n+2}{4} \right]} f \right|_{x \in S} = 0.
\label{f0}
\end{equation}
Let us denote the class of such functions by $\mathbf{F_2}$.

We shall also assume that function $r(t,\tau)$ is defined and is continuous on the set $D = \{ (t,\tau): (t,\tau) \in [0,T]\times [0,\infty) \}$ and $2\pi$-periodic in $\tau$. As above let represent it as the sum:
\begin{equation*}
r (t,\tau) = r_0 (t) + r_1 (t,\tau),
\end{equation*}
where $r_0$ is slow component and $r_1$ is oscillating component. Let us assume that $r_0 \in C ([0,T])$, $r_1 \in C (D).$

\begin{theorem}
The following asymptotic formula holds
\begin{equation}
\bigl\|u_{\omega} - u_0 \bigr\|_{C(\overline{\Pi})} = o(1), \; \omega \to \infty,
\label{thdirprob2}
\end{equation}
where $u_{\omega}$ is solution of problem \eqref{s1}-\eqref{s3}.
\label{thdirprob}
\end{theorem}

\noindent\textbf{The inverse problem 2}

Consider the problem \eqref{s1}--\eqref{s3} in domain $\Omega$ with boundary $S \in C^{2 \left[\frac{n}{2}\right] + 4}$.
Let coefficients of expression $L$ belong to the following Holder classes:
\begin{equation}
a_{ij} \in C^{3\left[\frac{n}{2}\right] + 6} (\overline{\Omega}), c \in C^{3\left[\frac{n}{2}\right] + 5} (\overline{\Omega}), \text{ãäå } \alpha \in (0,1), c (x) \geq 0, x \in \Omega.
\label{coefl2}
\end{equation}
We shall assume that function $r(t, \tau)$ is known, satisfies the Theorem \ref{thdirprob} conditions, and, moreover, $r_0 \in C^{1} ([0,T])$. Suppose there exist a point $t_0 \in (0,T]$ such that
\begin{equation}
|r_0 (t_0)| > |r_0 (0)|.
\label{r0t0}
\end{equation}
Let $\mathbf{R_2}$ denote the class of functions $r$ satisfy conditions above. We assume that function $f$ is unknown and belong to the class $\mathbf{F_2}$ .

Following lemm holds, where
\begin{equation*}
\Lambda_m (t) \equiv \int_0^t r_0 (s) \sin{\sqrt{\lambda_m} (t-s)} ds, t \in [0,T].
\end{equation*}

\begin{lemm}
For any function $r \in \mathbf{R_2}$ there exist values $c_0 > 0$ and $m_0 \in \mathbb{N}$ such that for every number $m \geq m_0$ we have $\Lambda_m (t_0) > \frac{c_0}{\lambda_m}$.
\label{lemminvprob1}
\end{lemm}
For brevity, we shall asuume that set $M_0 \equiv \{m : \Lambda_m (t_0) = 0\} = \emptyset$.

Concerning the system \eqref{s1}--\eqref{s3} with unknown function $f$, we supplement the problem with function $\psi$ such that
\begin{equation}
\psi \in C^{3\left[ \frac{n}{2} \right] + 7} (\overline{\Omega}), \left. \psi \right|_{x \in S} = \left. L \psi \right|_{x \in S} = \left. L^2 \psi \right|_{x \in S} = ... = \left. L^{ 3 \left[ \frac{n}{4} \right]+3} \psi \right|_{x \in S} = 0.
\label{psi}
\end{equation}

The inverse problem 2 is to find function $f\in \mathbf{F_2}$ for which the solution $u_{\omega} (x,t)$ of problem \eqref{s1}--\eqref{s3} satisfies the condition:
\begin{equation}
\bigl\| u_{\omega}(x, t_0) - \psi(x) \bigr\|_{C([0,\pi])} = o(1), \omega \to \infty.
\label{eq2.4}
\end{equation}

\begin{theorem}
Let functions $r_0, \psi$ and point $t_0$ satisfying the conditions specified above. Then inverse problem 2 uniquely solvable. At the same time, the function $f(x)$ calculated by
$f(x) = \sum_{m = 1}^{\infty} f_m y_m (x), \; f_m = \frac{\psi_m}{\Lambda_m}$.
\label{thinvprob1}
\end{theorem}

\subsection{The inverse problem 3}
\label{invproblem3}

In this section we consider again problem \eqref{s1}-\eqref{s3}. Assume that coefficients of operator $L$ satisfying to conditions \eqref{coefl2}, domain boundary $S \in C^{2 \left[\frac{n}{2}\right] + 4}$.

Let function $f$ and $r$ belong to $\mathbf{F_3}$ and $\mathbf{R_3}$ respectively:

$\mathbf{F_3}: f, Lf \in C^{\left[\frac{n}{2}\right] + 2} (\overline{\Omega}), \left. f \right|_{x \in S} = \left. L f \right|_{x \in S} = ... = \left. L^{\left[ \frac{n+6}{4} \right]} f \right|_{x \in S} = 0;$

$\mathbf{R_3}: r(t,\tau)$ is $2\pi$-periodic in $\tau$. As above let represent it as the sum:
\begin{equation*}
r (t,\tau) = r_0 (t) + r_1 (t,\tau),
\end{equation*}
where
\begin{equation*}
r_0 \in C^1 ([0,T]); r_1, r_{1t}, r_{1tt}, r_{1ttt} \in C(D).
\end{equation*}

We shall assume that function $r_0$ is known, and functions $f$ and $r_1$ are unknown. For brevity, as in Sec. \ref{dirproblem} suppose that set $M_0 \equiv \{ m, \Lambda_m (t_0) = 0 \} = \emptyset$.
Choose a $2\pi$-periodic with zero mean in second variable $\chi (t,\tau), \chi \in C^{3,2} (D), D = [0,T] \times [0, \infty)$, and function $\psi \in C^{3\left[ \frac{n}{2} \right] + 9} (\overline{\Omega})$ satisfying the conditions
\begin{equation}
\left. \psi \right|_{x \in S} = \left. L \psi \right|_{x \in S} = \left. L^2 \psi \right|_{x \in S} = ... = \left. L^{ 3 \left[ \frac{n}{4} \right]+4} \psi \right|_{x \in S} = 0.
\label{psi1}
\end{equation}
And let $x^{0} \in \Omega$ is a point at which $\widetilde{f}(x^{0}) \neq 0$, where
\begin{equation}
\widetilde{f} (x) = \sum_{m = 1}^{\infty} \widetilde{f}_m y_m (x), \; \widetilde{f}_m = \frac{\psi_m}{\Lambda_m}
\label{fw}
\end{equation}

Consider the functions $\varphi_0 (t), \varphi_1 (t), \varphi_2 (t)$, defined as follows. Function $\varphi_0 (t)$ is solution of Cauchy problem
\begin{equation}
\left\{\begin{array}{c}
\varphi_0'' (t) = \widetilde{f}(x^{0})r_0(t) + \int_0^t K(t,s) r_0(s) \, ds, \\
\varphi_0(0)=\varphi'_0(0)=0,
\end{array}\right.
\label{eq3.2}
\end{equation}
where
\begin{equation*}
K (t,s) = - \sum\limits_{m=1}^{\infty} \sqrt{\lambda_m} \widetilde{f}_m \sin{\sqrt{\lambda_m}(t-s)} y_m (x^{0}).
\end{equation*}
Functions $\varphi_1, \varphi_2$ satisfying the conditions
\begin{equation}
\varphi_1 (t) = \sum_{m = 1}^{\infty} \frac{\widetilde{b}_{1,m}}{\sqrt{\lambda_m}} y_m (x^{0}) \sin{\sqrt{\lambda_m} t},
\label{eq3.31}
\end{equation}
\begin{equation}
\varphi_2 (t) = \sum_{m = 1}^{\infty} y_m (x^{0}) \left( \widetilde{d}_m \cos{\sqrt{\lambda_m} t} + \frac{\widetilde{b}_{2,m}}{\sqrt{\lambda_m}} \sin{\sqrt{\lambda_m} t} \right),
\label{eq3.3}
\end{equation}
where
\begin{equation}
\widetilde{b}_{1,m} = -  \rho_{0\tau} (0,0) \widetilde{f}_m,
\label{b1}
\end{equation}
\begin{equation}
\widetilde{d}_m = - \rho_0 (0,0) \widetilde{f}_m,
\label{dm}
\end{equation}
\begin{equation}
\widetilde{b}_{2,m} = - ( 2 \rho_{1t} (0,0) + \rho_{0t} (0,0)) \widetilde{f}_m.
\label{b2}
\end{equation}

The inverse problem 3 is to find a functions $f$ and $r_1$ such that $f \in \mathbf{F_3}, r_1$ is $2\pi$-periodic in $\tau$ and, moreover, $r_1, r_{1t}, r_{1tt}, r_{1ttt} \in C(D)$ for which the solution $u_{\omega} (x,t)$ of problem \eqref{s1}-\eqref{s3} satisfies the conditions
\begin{equation}
\left\| u_{\omega} (x^{0},t) - \left[ \varphi_0 (t) + \frac{1}{\omega}\varphi_1 (t) + \frac{1}{\omega^2} \bigl( \varphi_2 (t) + \chi (t,\omega t) \bigr) \right] \right\|_{C([0,T])} = o (\omega^{-2}),
\label{inv21}
\end{equation}
\begin{equation}
\bigl\| u_{\omega}(x, t_0) - \psi(x) \bigr\|_{C(\overline{\Omega})} = o(1), \omega \to \infty.
\label{inv22}
\end{equation}

\begin{theorem}
Let functions $r_0, \psi, \chi$ and points $x^{0}, t_0$ satisfying the conditions specified above. Then inverse problem 3 uniquely solvable. At the same time, the function $f(x) = \widetilde{f}(x)$ calculated by \eqref{fw}, and
\begin{equation}
r_1 (t,\tau) = (f(x^{0}))^{-1} \frac{\partial^2}{\partial \tau^2} \chi (t,\tau).
\label{eqr1}
\end{equation}
\label{thinvprob2}
\end{theorem}

\section{Proof of the main results}

\noindent\textbf{Proof of the Theorem \ref{thdirprob}.} \newline

Consider the function
\begin{equation}
W_{\omega}(x,t) = u_{\omega}(x,t) - u_0(x,t) = \sum_{m = 1}^{\infty} \frac{f_m y_m(x)}{\lambda_m} \int\limits_0^t \sin{\sqrt{\lambda_m}(t-s)} r_1 (s, \omega s) ds,
\label{wsum}
\end{equation}
Note that, in view of Lemmas 1,2 and Cauchy–Schwarz inequality, the series in right-hand side of \eqref{wsum} converges uniformly with respect to $t \in [0,T]$. Represent $W_{\omega}(x,t)$ in the form
\begin{multline*}
W_{\omega}(x,t) = \sum_{m = 1}^{m_0} \frac{f_m y_m(x)}{\lambda_m} \int\limits_0^t \sin{\sqrt{\lambda_m}(t-s)} r_1 (s, \omega s) ds +
\\ \sum_{m = m_0 + 1}^{\infty} \frac{f_m y_m(x)}{\lambda_m} \int\limits_0^t \sin{\sqrt{\lambda_m}(t-s)} r_1 (s, \omega s) ds \equiv S_{\omega,1} + S_{\omega,2}, m_0 \in \mathbb{N}.
\end{multline*}
Let $\varepsilon$ is arbitrary value. Taking into account uniform convergence of the series \eqref{wsum}, we take number $m_0$ sufficiently large such that for all $m, m \geq m_0,$ and $\omega > 0$
\begin{equation}
\| S_{\omega, 2} \|_{C (\overline{\Omega})} < \frac{\varepsilon}{2}.
\label{eq24}
\end{equation}

For the estimation of $ S_{\omega, 1}$ choose $\delta > 0$ so small that
\begin{equation}
\int\limits_0^{\delta} \sin{\sqrt{\lambda_m}(t-s)} r_1 (s, \omega s) ds < \frac{\varepsilon}{2 m_0 s_0},
\end{equation}
where $s_0 = \max_{1 \leq i \leq m_0} \left| \frac{f_i}{\lambda_i} \right| \| y_i \|_{C(\overline{\Omega})}$.
Further, considering $t \in (s,T]$, we divide the interval $[\delta,t]$ into $k$ equal parts $[t_j,t_{j+1}), j=\overline{0,k-1},$ and apply the relation
\begin{multline*}
\int\limits_{\delta}^t \sin{\sqrt{\lambda_m}(t-s)} r_1 (s, \omega s) ds = \\
 \sum_{j = 0}^{k-1} \left[ \int\limits_{t_j}^{t_{j+1}} \sin{\sqrt{\lambda_m}(t-s)} r_1 (s, \omega s) ds - \int\limits_{t_j}^{t_{j+1}} \sin{\sqrt{\lambda_m}(t-t_j)} r_1 (t_j, \omega s) ds \right] + \\
   \sum_{j = 0}^{k-1} \int\limits_{t_j}^{t_{j+1}}\sin{\sqrt{\lambda_m}(t-t_j)} r_1 (t_j, \omega s) ds = S_1 + S_2.
\end{multline*}
Choose $k = k(t)$ so large that
\begin{equation}
| S_1 | < \frac{\varepsilon}{4 m_0 s_0}
\label{eq25}
\end{equation}
for all $m: m < m_0$ and $\omega > 0$.

Further, in view of equality $\left\langle r_1 (t,\tau) \right\rangle_{\tau} = 0$, we choose $\omega_0$ sufficiently large that
\begin{equation}
| S_{2} | < \frac{\varepsilon}{4 m_0 s_0}
\label{eqth3}
\end{equation}
for given $k, t \in [0,T],$ and any $\omega > \omega_0$.

Since inequalities \eqref{eq25}, \eqref{eqth3} there exist number $\omega_0 > 0$ such that
\begin{equation}
| S_{\omega,1} | < \frac{\varepsilon}{2}
\label{eq27}
\end{equation}
for any $\omega > \omega_0$. Relations \eqref{eq24}, \eqref{eq27} imply the  relation \eqref{thdirprob2}. This completes the proof of Theorem \ref{thdirprob}.

\noindent\textbf{Proof of the Lemma \ref{lemminvprob1}.} \newline
Choose $t_0$ that $|r_0 (t_0)| > |r_0 (0)|$ and apply the relation
\begin{equation*}
\Lambda_m (t_0) = \int\limits_0^{t_0} \frac{\sin \sqrt{\lambda_m}(t_0 - s)}{\sqrt{\lambda_m}} r_0 (s) ds = \frac{r_0 (t_0) - r_0 (0) \cos \sqrt{\lambda_m}t_0}{ \lambda_m} +  \int\limits_0^{t_0} \frac{\cos \sqrt{\lambda_m}(t_0-s)}{ \lambda_m} r_0' (s) ds.
\end{equation*}
Taking into account the condition \eqref{r0t0}, note that $|r_0 (t_0)| \neq |r_0 (0) \cos \sqrt{\lambda_m} t_0|$ for all $m \in \mathbb{N}$. Thus there exist positive values $c_0$ and $m_0$ such that
\begin{equation*}
|\Lambda_m (t_0)| > \frac{c_0}{\lambda_m}
\end{equation*}
for $m > m_0$. The Lemma is proved.

\noindent\textbf{Proof of the Theorem \ref{thinvprob1}.} \newline
Choose $t_0$ that $|r_0 (t_0)| > |r_0 (0)|$. We assume that the function $f \in \mathbf{F_2}$ is found. It follows from Theorem \ref{thdirprob} and conditions \eqref{eq2.11}, \eqref{eq2.4} that
\begin{equation*}
\sum_{m=1}^{\infty} f_m y_m (x) \Lambda_m (t_0)  = \sum_{m=1}^{\infty} \psi_m y_m (x).
\end{equation*}
For $\Lambda_m \neq 0, m \in \mathbb{N} (M_0 = \emptyset)$ we obtain
\begin{equation*}
f(x) = \sum_{m=1}^{\infty} f_m y_m (x), f_m = \frac{\psi_m}{\Lambda_m (t_0)}.
\end{equation*}
It remains to show that $f$ belongs to class $\mathbf{F_2}$.

In the first place we shall show that function $f \in C^{\left[\frac{n}{2}\right]+2} (\overline{\Omega})$. Let us consider the series
\begin{multline*}
L^{\left[\frac{n}{2}\right]+2} f(x) = \sum_{m=1}^{\infty} \frac{\psi_m}{\Lambda_m (t_0)} L^{\left[\frac{n}{2}\right]+2} y_m (x) = \\
\sum_{m=1}^{m_0} \frac{\psi_m}{\Lambda_m (t_0)} L^{\left[\frac{n}{2}\right]+2} y_m (x) + \sum_{m=m_0+1}^{\infty} \frac{\psi_m}{\Lambda_m (t_0)} \lambda_m^{\left[\frac{n}{2}\right]+2} y_m (x) = Y_1 + Y_2,
\label{fxsum}
\end{multline*}
In view of Lemmas \ref{ilinlemm0}, \ref{ilinlemm}, \ref{lemminvprob1} and Cauchy–Schwarz inequality, series $Y_2$ may be estimate as follows
\begin{equation*}
\| Y_2 \|_{L_2 (\Omega)} \leq \frac{1}{c_0} \sqrt{ \sum_{m = m_0+1}^{\infty} \frac{y^2_m (x)}{\lambda_m^{\left[ \frac{n}{2} \right] + 1}} \cdot \sum_{m = m_0+1}^{\infty} \psi_m^2 \lambda_m^{3\left[ \frac{n}{2} \right] + 7} },
\end{equation*}
where $c_0$ and $m_0$ are the same as in Lemma \ref{lemminvprob1}.

Further, let $g$ denote the function $g (x) = L^{\left[\frac{n}{2}\right]+2} f(x), g \in L_2 (\Omega)$. Thus
\begin{equation*}
f = L^{- \left[\frac{n}{2}\right]-2} g.
\end{equation*}
As in the Lemma \ref{th5} consider $D^{\left[\frac{n}{2}\right]+2}$ is the derivative of order $\left[\frac{n}{2}\right]+2$, and then apply it to the function $f$, we obtain
\begin{equation*}
D^{\left[\frac{n}{2}\right]+2} f = D^{\left[\frac{n}{2}\right]+2} L^{- \left[\frac{n}{2}\right]-2} g.
\end{equation*}
From the Lemma \ref{th5} it follows that function $D^{\left[\frac{n}{2}\right]+2} f$ is continuous.

Note that, since proved smoothness of the function $f$ and properties of the eigenfunctions $y_m (x)$ it follows that for founded function $f(x)$ conditions $\eqref{f0}$ are hold. This completes the proof of Theorem \ref{thinvprob1}.

\noindent\textbf{Proof of the Theorem \ref{thinvprob2}.} \newline
Let the hypotheses of current theorem holds. Then according to Theorem \ref{thinvprob1} the inverse problem 2 with given functions $r_0, \psi$ and point $t_0$ is uniquely solvable, and function $\widetilde{f}$ calculable by \eqref{fw} is the inverse problem 2 solution.
Providing similar to Theorem \ref{thinvprob1} reasoning we obtain that $f \in \mathbf{F_3}$.

Further, consider system \eqref{s1}-\eqref{s3} with $f(x,t) = \widetilde{f}(x)$, and also the inverse problem 1 with given functions $\chi, \varphi_i, i=\overline{0,2}$ and point $x^0$. In view condition \eqref{eq3.2}, the function $r_0 (t)$ satisfies Volterra equation of the second kind
\begin{equation*}
\varphi_0'' (t) = \widetilde{f}(x^{0})r_0(t) + \int_0^t K(t,s) r_0(s) \, ds,
\end{equation*}
\begin{equation*}
K (t,s) = - \sum\limits_{m=1}^{\infty} \sqrt{\lambda_m} \widetilde{f}_m \sin{\sqrt{\lambda_m}(t-s)} y_m (x^{0}).
\end{equation*}

From theorem \ref{babich-t4} it follows that the inverse problem 1 with given data is uniquely solvable, moreover, its solution may be represented in form $r (t,\tau) = r_0 (t) +r_1 (t,\tau)$, where $r_1$ calculated by \eqref{eqr1}. Because of the conditions on function $r_0$ the inverse problem 1 solution $r$ belongs to the class $\mathbf{R_3}$.

Hence pair of functions $\widetilde{f}, r_1$ is solution of the inverse problem 3. This completes the proof of this Theorem.

Valeriy Borisovich Levenshtam - Dr. Phys.-Math.Sci., Prof.
Southern Federal University, Rostov-on-Don, 334006 Russia
Southern Mathematical Institute, Vladikavkaz Scientific Center,
Russian Academy of Sciences, Vladikavkaz, 362027 Russia

Pavel Vasil'evich Babich - aspirant
Southern Federal University, Rostov-on-Don, 334006 Russia


\newpage

\begin{thebibliography}{text}

\bibitem{bib3}
{\it M. M. Lavret'ev, K. G. Reznitskaya, and V. G. Yakhno} One-Dimensional Inverse Problems of Mathematical Physics (Nauka, Novosibirsk, 1982) [in Russian].

\bibitem{bib4}
{\it V. G. Romanov} Inverse Problems of Mathematical Physics  (Nauka, Moscow, 1984) [in Russian]

\bibitem{bib5}
{\it A. M. Denisov} Introduction to the Theory of Inverse Problems (Nauka, Moscow, 1994) [in Russian].

\bibitem{bib6}
{\it Anikonov Jn.~E.} Multidimentional inverse and Ill-posed problems for diffentional equations (Utrecht, VSP, 1995)

\bibitem{bib7}
{\it Anikonov Jn.~E.} Formulas in inverse and Ill-posed problems (Utrecht, VSP, 1997)

\bibitem{bib8}
{\it Anikonov Jn.~E., Bubhov B.A., Erokhin G.N.} Inverse and Ill-posed source problems. (Utrecht, VSP, 1997)

\bibitem{bib9}
{\it Prilepko A.~I., Orlovsky D.~G., Vasin I.~A.} Methods for solving inverse problems in mathematical physics (N.Y.-Basel, Marcel Dekker, Inc., 1999)

\bibitem{bib10}
{\it Anikonov Jn.~E.} Inverse problems for kinetic and other evolution equations (Utrecht, VSP, 2001)

\bibitem{bib11}
{\it Belov Jn.~Ya.} Inverse problems for partial differential equation (Utrecht, VSP, 2002)

\bibitem{bib12}
{\it Lavrentiev M.~M.} Inverse problems of mathematical physics (Utrecht, VSP, 2003)

\bibitem{bib13}
{\it Megrabov A.~G.} Forvard and inverse problems for Hyperbolic, elliptic and mixed type equations (Utrecht, VSP, 2003)

\bibitem{bib14}
{\it Ivanchov M.} Inverse problems for equations of parabolic type  (VNTL, Publishers, 2003)

\bibitem{bib15}
{\it V. G. Romanov} Stability in Inverse Problems (Nauchnyi Mir, Moscow, 2005) [in Russian]

\bibitem{bib16}
{\it S.I. Kabanikhin} Inverse and Ill-Posed Problems (Sib. Nauchn. Izd., Novosibirsk, 2008) [in Russian].

\bibitem{bib17}
{\it A. M. Denisov} "Asymptotic expansions of solutions to inverse problems for a hyperbolic equation with a small parameter multiplying the highest derivative", Zh. Vychisl. Mat. Mat. Fiz. 53 (5), 744-752 (2013) [Comput. Math. Math. Phys. 53 (5), 580-587 (2013)].

\bibitem{bib171}
{\it V. L. Kamynin} "Inverse problem of simultaneously determining the right-hand side and the coefficient of a lower order derivative for a parabolic equation on the plane", Differ. Uravn. 50 (6), 795-806 (2014) [Differ. Equations 50 (6), 792-804 (2014)].

\bibitem{bib18}
{\it A. M. Denisov} "Problems of determining the unknown source in parabolic and hyperbolic equations," Zh. Vychisl. Mat. Mat. Fiz. 55 (5), 830-835 (2015) [Comput. Math. Math. Phys. 55 (5), 829–833 (2015)].

\bibitem{bib211}
{\it P. V. Babich and V. B. Levenshtam} "Recovery of a rapidly oscillating absolute term
in the multidimensional hyperbolic equation" // Mathematical Notes, Vol. 104, No. 4, 489-497 (2018).


\bibitem{bib19}
{\it S. M. Zen'kovskaya and I. B. Simonenko} "On the influence of a high-frequency vibration on the origin of convection," Izv. Akad. Nauk SSSR Ser. Mekh. Zhidk. Gaza, No. 5, 51-55 (1966).

\bibitem{bib20}
{\it  I. B. Simonenko} "A justification of the averaging method for a problem of convection in a field of rapidly oscillating forces and for other parabolic equations," Mat. Sb. 87 (129) (2), 236-253 (1972) [Math. USSR-Sb. 16 (2), 245-263 (1972)].

\bibitem{bib21}
{\it V. B. Levenshtam} "The averaging method in the convection problem with high-frequency oblique vibrations," Sibirsk. Mat. Zh. 37 (5), 1103-1116 (1996) [Sib. Math. J. 37 (5), 970-982 (1996)].

\bibitem{bib22}
{\it V. B. Levenshtam}  "Asymptotic expansion of the solution to the problem of vibrational convection," Zh. Vychisl. Mat. Mat. Fiz. 40 (9), 1416-1424 (2000) [Comput. Math. Math. Phys. 40 (9), 1357-1365 (2000)]

\bibitem{bib221}
{\it V. B. Levenshtam} "Asymptotic integration of a problem of convection" Sibirsk. Mat. Zh. 30 (4), 554-559 (1989) [Sibirsk. Mat. Zh. 30(4), 69-75 (1989)]

\bibitem{bib2211}
{\it V. B. Levenshtam} "Justification of the averaging method for the convection problem with high-frequency vibrations" Sibirsk. Mat. Zh. 34 (2), 280-296 (1993) [Sibirsk. Mat. Zh. 34 (2), 92-109 (1993)]


\bibitem{bib1}
{\it P.V. Babich, V.B. Levenshtam} "Direct and inverse asymptotic problems high-frequency terms" Asymptotic Analysis. 97, 329-336 (2016)

\bibitem{bib2}
{\it P. V. BabichEmail authorV. B. LevenshtamS. P. Prika} "Recovery of a Rapidly Oscillating Source in the Heat Equation from Solution Asymptotics"  Comput. Math. and Mathematical Phys. 57(12), 1908-1918 (2017) [57(12), 1955-1965 (2017)]


\bibitem{bib23}
{\it V. A. Il'in}  "The solvability of mixed problems for hyperbolic and parabolic equations," Uspekhi Mat. Nauk 15 (2 (92)), 97-154 (1960) [Russian Math. Surveys 15 (1), 85-142 (1960)]

\bibitem{bib24}
{\it M.A. Krasnoselskii, P.P. Zabreiko, E.I. Pustylnik, P.E. Sobolevskii} "Integral operators in spaces of summable functions" (Noordhoff International Publishing, Leyden, 1976)

\bibitem{bib29}
{\it S. Agmon, A. Douglis, L. Nirenberg} Estimates near the boundary for solutions of elliptic partial differential equations satisfying general boundary conditions. ( New York, Interscience Publishers, 1959)

\end{thebibliography}
\end{document}